\documentclass{ws-procs9x6}

\newcommand{\be}{\begin{equation}}
\newcommand{\ee}{\end{equation}}

\newcommand{\tE}{\tilde{\mathcal{E}}}

\newcommand{\mjsum}{\sum_{m,j=0}^\infty \!\!{}^{{}^\prime}}
\newcommand{\im}{\mathrm{Im}}
\newcommand{\re}{\mathrm{Re}}

\newcommand{\jsum}{\sum_{j=0}^\infty \!{}^{^\prime}}

\newcommand{\zint}{\int_0^\infty}
\newcommand{\um}{\underline{m}}

\newcommand{\Em}{\mathcal{E}_m}

\newcommand{\lsx}{\lambda^2_{mp}(x)}

\begin{document}

\title{ELECTROMAGNETIC CASIMIR EFFECT IN WEDGE GEOMETRY AND THE 
ENERGY-MOMENTUM TENSOR IN MEDIA}

\author{I. BREVIK$^*$ and S. {\AA}. ELLINGSEN$^\dag$}
\address{Department of Energy and Process Engineering, Norwegian University of 
Science and Technology, N-7491 Trondheim, Norway\\
$^*$E-mail: iver.h.brevik@ntnu.no\\
$^\dag$E-mail: simen.a.ellingsen@ntnu.no}
\author{K. A. MILTON}
\address{Oklahoma Center for High Energy Physics and H.L.~Dodge Department of
Physics and Astronomy, The University of Oklahoma, Norman, OK 73019, USA\\
E-mail: milton@nhn.ou.edu}

\begin{abstract}
The wedge geometry closed by a circular-cylindrical arc
is a nontrivial generalization of the cylinder,
which may have various applications.  If the
radial boundaries are not perfect conductors, the angular eigenvalues
are only implicitly determined.  When the speed of light is the
same on both sides of the wedge, the Casimir energy is
finite, unlike the case of a perfect conductor, where there
is a divergence associated with the corners where the radial planes meet
the circular arc.  We advance the study of this system by reporting
results on the temperature dependence for the conducting situation.
We also discuss the appropriate choice of the electromagnetic
energy-momentum tensor.
\end{abstract}

\section{Introduction}

Casimir theory for the wedge geometry continues to attract
interest. The reasons for this are many-faceted -- probably the
most important one being that the material boundaries are plane,
thus avoiding some of the formal divergences that so often plague
calculations in the presence of curved boundaries. The wedge
geometry moreover implies a formalism closely related to that of
cylindrical geometry, and actually also to that of cosmic string
theory. Finally, the wedge geometry is a convenient testing ground
for experimental tests of Casimir-Polder forces. 

The Casimir energy and
 stress in a wedge geometry was approached already in the 1970s
 \cite{dowker78,deutsch79}. Various embodiments of the wedge with
 perfectly conducting walls were treated by Brevik and co-workers
 \cite{brevik96,brevik98,brevik01} and others
 \cite{nesterenko02,razmi05}. More recently a wedge intercut by a
 cylindrical shell was considered by Nesterenko and 
co-workers\cite{nesterenko01,nesterenko03}. Local Casimir
 stresses were considered by Saharian and co-workers
 \cite{rezaeian02,saharian07,saharian08}. The interaction of an
 atom with a wedge was studied experimentally by Sukenik {\it et al.}
 \cite{sukenik93}. The theory of that interaction was worked out
 by Barton \cite{barton87} and others 
\cite{skipsey05,skipsey06,mendes08,rosa08}. The semitransparent 
wedge has very recently been considered
 by Milton, Wagner,  and Kirsten\cite{wagner10}.
 The closely related case of circular symmetry has been treated in
 several papers, dealing with a perfectly conducting circular
 boundary \cite{deraad81,gosdzinsky98,lambiase99}, as
 well as the case of a dielectric circular boundary
 \cite{cavero05,cavero06,romeo05,brevik07}.

\begin{figure}
  \begin{center}
  \psfig{file=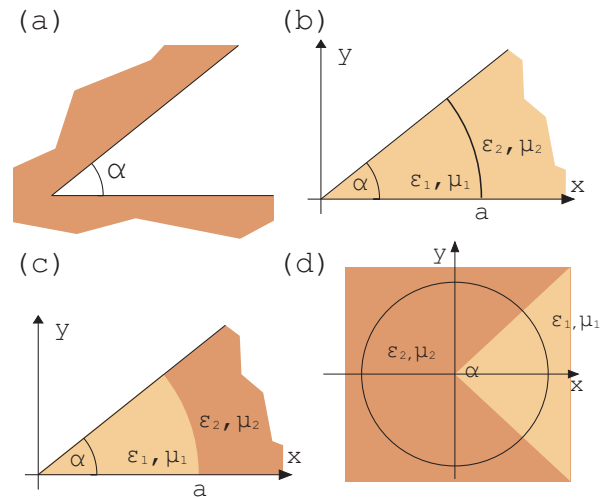, width = 3.2in}
  \end{center}
  \caption{Wedge geometries. (a) The perfectly conducting wedge geometry. 
(b) The geometry of a wedge intercut by a perfectly conducting cylindrical arc.
 (c) Wedge with magnetodielectric arc. (d) Diaphanous wedge in a perfectly 
conducting cylindrical shell.}
  \label{fig:wedge}
\end{figure}

The typical wedge
geometry is sketched in figure \ref{fig:wedge}a. The planes are situated at
$\theta=0$ and $\theta=\alpha$. We shall assume in the present
paper that the interior of the wedge is filled with an isotropic
medium of spatially constant and nondispersive refractive index 
$n=\sqrt{\varepsilon\mu}$.

In the simplest version of the wedge model, the planes are taken
to be perfectly conducting. Various modifications of this simple
wedge model can be envisaged. In sections 3 and 4 below we
consider two generalizations of the simple wedge geometry, called
Wedge I and II, in which the interior region is closed by a
circular boundary thus implying an eigenvalue problem for the
photon frequencies. The Wedge II model treated in section 4, in
particular, removes the strict perfect boundary condition of the
radial walls.  The material of these two sections is based on two recent
papers \cite{brevik09,ellingsen09}. New developments are a closer
examination of the behavior at finite temperature.
As an introductory step, we delineate in the next section the
essentials of classic Casimir theory for the perfectly conducting
wedge. In section 6 we relate the macroscopic electromagnetic
theory for the dielectric wedge region to  the more  general
question about which electromagnetic energy-momentum tensor is to
be preferred in media. This one-hundred-year-old question has actually
attracted considerable interest recently.

 \section{Extracts from the Classic Theory for the  Perfectly Conducting Wedge}

The geometry is shown in Fig.~\ref{fig:wedge}a and was considered in a number 
of publications\cite{dowker78,deutsch79,brevik96,brevik98,brevik01}. The 
governing equation for the
Fourier transform of Green's function ${\bf \Gamma}(x,x')$ is
\begin{equation}
 {\bf \nabla \times \nabla \times
\Gamma(r,r'},\omega)-\varepsilon \mu \omega^2{\bf
\Gamma(r,r'},\omega)=-\mu \omega^2{\bf 1}\delta({\bf r-r'}).
\label{1}
\end{equation}
After solving this equation in terms of the scalar Green's
functions $F_m(r,r')$ and $G_m(r,r')$ \cite{brevik01} we can
calculate the effective field products for the electric fields as
$i\langle E_i({\bf r})E_k({\bf r'}\rangle_\omega=\Gamma_{ik}({\bf
r,r'},\omega)$. 
The corresponding products for the magnetic fields follow from
Maxwell's equations. The points $\bf r$ and $\bf r'$ are assumed
to be close but not coincident.

The effective products can now be inserted in the electromagnetic
energy-momentum tensor $S_{\mu\nu}$, whose spatial part in
classical notation is, in Heaviside-Lorentz units,
\begin{equation}
S_{ik}=-E_iD_k-H_iB_k+\frac{1}{2}\delta_{ik}({\bf E\cdot D+H\cdot
B}). \label{3}
\end{equation}
A lengthy calculation leads to the expression (Minkowski metric
assumed)
\begin{equation}
\langle S_{\mu\nu}({\bf r})\rangle
 =\frac{1}{720\pi^2n}\frac{1}{r^4}\left(\frac{\pi^2}
 {\alpha^2}+11\right)\left(\frac{\pi^2}
 {\alpha^2}-1\right)\rm{diag}(1,-3,1,1), \label{4}
 \end{equation}
 where we have  subtracted off the
 term $\langle
S_{\mu\nu}^{\alpha=\pi}({\bf r})\rangle$ corresponding to a plane
sheet. The components are ordered as $\langle
S_{\mu\nu}\rangle=\langle
S_{rr},S_{\theta\theta},S_{zz},-w\rangle$, where $w$ is the energy
density.

The expression (\ref{4}) refers to zero temperature. It is worth
noticing that in the limit $\alpha \rightarrow 0, r\rightarrow
\infty$ such that $\alpha r$ becomes the separation between
parallel plates, the expression agrees with Barton
\cite{barton87}. Moreover, the expression agrees with that of a
cosmic string if the string's deficit angle $8\pi GM$ is
identified with $2\pi-2\alpha$ \cite{brevik09}.

\section{ Wedge I: Perfectly Conducting Walls;  Circular Boundary at $r=a$}

The geometry is shown in Fig.~\ref{fig:wedge}b and \ref{fig:wedge}c. 
The walls are
perfectly conducting as before, while we assume now  that there is
as boundary a circular arc with radius $a$. The wedge thus has an
interior region $r<a$ (region 1) where the material parameters are
$\varepsilon_1$ and $\mu_1$, and an exterior region $r>a$ (region
2) with analogous parameters $\varepsilon_2$ and $\mu_2$. The
materials are assumed to be nondispersive. We assume the
diaphanous (isorefractive) condition 
$\varepsilon_1\mu_1=\varepsilon_2\mu_2=n^2$. 
The transverse wave numbers $k_\perp$ in the two regions are
accordingly the same,
$k_\perp^2=n^2\omega^2-k_z^2$. 
It is convenient to introduce the symbol
$p=\pi/\alpha$, 
and also
$\lambda_\nu(x)=(I_\nu(x) K_\nu(x))'$,
where $I_\nu$ and $K_\nu$ are modified Bessel functions.

\subsection{The boundary $r=a$ being perfectly conducting}

This is the simplest case. Detailed expansions of the electric and
magnetic fields are given in Ref.~\refcite{brevik09}. In region 1
there are two independent polarizations, one TM polarization where
the mode eigenvalues are determined by $J_{mp}(k_\perp a)=0$ with
$m=1,2,3,...,$ and one TE polarization where the eigenvalues are
determined by $J_{mp}'(k_\perp a)=0$.
 One azimuthally symmetric TE
mode exists, corresponding to $m=0$, but there is no such TM mode.

In region 2 the TM polarization yields $H_{mp}^{(1)}(k_\perp
a)=0$, $m=1,2,3,..$, whereas the TE polarization yields
${H_{mp}^{(1)}}'(k_\perp a)=0$, $m=0,1,2,\dots$. Summing over all
modes and making use of the argument principle, we arrive at the
following expression for the total zero-point energy
\begin{equation}
\tilde{\mathcal{E}}= \frac1{4\pi n
a^2}\sum_{m=0}^\infty{}'\int_0^\infty
 dx \,x\ln[1-x^2\lambda^2_{mp}],\label{8}
\end{equation}
the prime meaning that the mode $m=0$ is counted with half weight.
This is the boundary-induced contribution to the zero-point
energy. If the boundary $r=a$ were removed and either the interior
or the exterior medium were to fill the whole region, we would get
$\tilde{\mathcal{E}}=0.$ Moreover, we have omitted a zero-mode
divergence caused by the sharp corners where the arc meets the
wedge. If $p=1$, Eq.~(\ref{8}) is one-half that for a conducting
circular cylinder.

\subsection{Dielectric boundary at $r=a$}

The most important change compared with the previous subsection is
that regions 1 and 2 become coupled via electromagnetic boundary
conditions. Assuming $n_1=n_2$ we find the following simple expression for
the zero-point energy
\begin{equation}
\tilde{\mathcal{E}}= \frac1{4\pi n
a^2}\sum_{m=0}^\infty{}'\int_0^\infty
 dx \,x\ln[1-\xi^2 x^2\lambda^2_{mp}],\label{9}
 \end{equation}
 with
 \begin{equation}
 \xi=\frac{\varepsilon_2-\varepsilon_1}{\varepsilon_2+\varepsilon_1}.
 \label{10}
 \end{equation}
 The conducting case is obtained by setting $\xi=1$. The case $n_1\neq n_2$ is 
more complicated, but for weak-coupling, $|\varepsilon_1-\varepsilon_2|\ll1$,
a self-energy can still be extracted\cite{brevik09} by 
generalizing the work done on dielectric cylinders\cite{cavero05, cavero06}.

The expressions (\ref{8}) and (\ref{9}) are still not in general
 finite. A finite self-energy can be extracted from this
 formula by a method of zeta function regularisation\cite{milton99}, 
 generalizing the standard formal result for a
 circular cylinder $(p=1)$ \cite{deraad81}. Further details and numerical 
results are reported in Ref.~\refcite{brevik09}.

\section{Wedge II: Diaphanous (Isorefractive) Wedge in cylindrical shell.}

Consider the geometry of Fig.~\ref{fig:wedge}d wherein a diaphanous 
magnetodielectric wedge ($n_1 = n_2$) inside a perfectly conducting cylindrical
shell of radius $a$ is considered. More details were published in 
Ref.~\refcite{ellingsen09}. The sum over orders of the Bessel function partial 
waves is now not simply equidistant values $\nu=mp$ as before but the zeros of 
the dispersion function for $\nu=i\eta$:
\be
  D(i\eta;\alpha) = \sinh^2\eta\pi - \xi^2 \sinh^2\eta(\pi-\alpha). 
\ee
The reflection coefficient $\xi$ was defined in Eq.~(\ref{10}). In the absence 
of any wedge this becomes $D_0(i\eta) = \sinh^2\eta\pi$. The energy of the 
diaphanous wedge enclosed by a perfectly conducting cylindrical shell is thus 
found as
\be
  \tE = \frac1{16\pi^3 i}\int_{-\infty}^\infty dk \int_{-\infty}^\infty d\zeta
\zeta \int_{-\infty}^\infty d\eta \left(\frac{D'}{D}-\frac{D_0'}{D_0}\right) 
\frac{d}{d\zeta}\ln [1-x^2\lambda_{i\eta}^2(x)].
\ee

In the non-dispersive case where $\xi$ is independent of $\zeta$ this may be 
simplified by means of partial integration w.r.t.~$\zeta$, introduction of 
polar coordinates, and use of symmetry properties, to
\be
  \tE =-\frac{1}{4\pi^2 na^2}\int_{0}^\infty d\eta \left(\frac{D'}{D}
-\frac{D_0'}{D_0}\right)\int_0^\infty dx\,x\, \im \ln [1-x^2\lambda_{i\eta}^2(x)].
\ee
Since $\re \{x^2\lambda^2_{i\eta}\}\leq 1$ we may use for numerical purposes
\[
  \im\ln(1-x^2\lambda^2_{i\eta}) = -\arctan\frac{x^2\im\lambda^2_{i\eta}}
{1-x^2\re\lambda^2_{i\eta}}.
\]

\begin{figure}
  \begin{center}
  \psfig{file=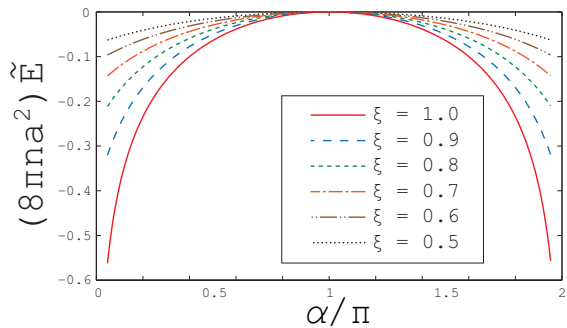, width = 3in}
  \end{center}
  \caption{The energy per length of the wedge, divided by a reference energy 
$(8\pi n a^2)^{-1}$, as a function of opening angle $\alpha$ for different 
values of the reflection coefficient $\xi$.}
  \label{fig:Eal}
\end{figure}

A numerical evaluation of this energy expression was performed in 
Ref.~\refcite{ellingsen09} and the details of the procedure will not be 
iterated here. The energy is plotted as a function of the opening angle 
$\alpha$ for different values of the reflection coefficient $|\xi|$ in 
Fig.~\ref{fig:Eal}.

\section{Considerations of finite temperature}

We present for the first time some considerations on the Casimir energy of a 
closed perfectly conducting
wedge  such as in figure \ref{fig:wedge}b when 
$T>0$. (For earlier work on the cylinder 
at high temperatures see Ref.~\refcite{bordag02}.)
 By letting $p=1$ in the following, these considerations automatically 
apply to the case of the perfectly conducting cylinder. We consider the finite 
part of the energy, given for zero $T$ in equation (\ref{8}). As is customary, 
finite temperature implies a compactification of the imaginary time axis so 
that the integral over imaginary frequencies $i\zeta$ becomes a sum over 
Matsubara frequencies $\zeta_j = 2\pi j T$:
\be
   \zint d\zeta f(\zeta) \to 2\pi T \jsum f(\zeta_j).
\ee
A little calculation shows that the expression (\ref{8}) may then be written 
$\tE=\sum_{m=0}^{\infty\prime} \Em$ where
\be
    \Em = \frac{T}{\pi a} \jsum \int_{j\tau}^\infty \frac{dx\,x}
{\sqrt{x^2-j^2 \tau^2}} \ln[1-x^2\lsx] =  \frac{T}{\pi a}\jsum e_{m,j}.
\ee
The nondimensional temperature is $\tau=2\pi naT$. As previously encountered 
at zero temperature the simple expression for $\tE$ is divergent and the 
challenge is to regularize it. We follow a scheme closely analogous to that of 
Ref.~\refcite{milton99} (c.f.~appendix A of Ref.~\refcite{brevik09}) using zeta
functions by subtracting and adding the asymptotic behaviour of the integrand,
\be
    \ln[1-x^2 \lsx] \sim -\frac{x^4}{4(m^2 p^2 + x^2)^3}, ~~m,x\to \infty.
\label{masymp}
\ee
This asymptotic behaviour is responsible for divergences both for $j\to\infty$ 
and $m\to \infty$. The work on this problem is still in progress and details of
the calculations and further discussion will be published 
elsewhere\cite{ellingsen10}. 

Adding and subtracting the asymptotic behaviour we write
\be\label{addsubtract}
  \Em = \frac{T}{\pi a} \jsum\left[ \tilde{e}_{\um,j} - \frac1{4}
\int_{j\tau}^\infty \frac{dx\, x^5}{\sqrt{x^2-j^2 \tau^2}
(\um^2 p^2 + x^2)^3}\right],
\ee
using compact notation $\um$ defined as $\um(m)=m$ for $m\geq 1$ and 
$\um(0)=1/p$. The symbol $\tilde{e}$ implies that the leading asymptotic term 
(\ref{masymp}) has been subtracted from the integrand, with $\um$ replacing 
$m$. The double sum resulting from the integral in (\ref{addsubtract}) is 
formally divergent but may be regularised by use of the Chowla-Selberg 
formula\cite{elizalde95}. After some calculation we obtain the following result
\begin{subequations}
\begin{align}
  \tE =& \frac{T}{\pi a} \mjsum\tilde{e}_{\um,j}-\frac{\tau(3-3\tau
\partial_\tau + \tau^2\partial^2_\tau)[1+2 \mathcal{K}(\tau) + \mathcal{S}
(\tau,p)]}{512\pi na^2};\\
  \mathcal{K}(\tau) =& \sum_{j=1}^\infty \frac{j\tau - \sqrt{1+j^2\tau^2}}
{j\tau \sqrt{1+j^2\tau^2}},\\
  \mathcal{S}(\tau,p) =& \frac{2}{p}[\gamma - \ln (4\pi p/\tau)] + 
\frac{8}{p}\sum_{l=1}^\infty \sigma_0(l)K_0(2\pi l \tau /p),
\end{align}
\end{subequations}
where $\gamma$ is Euler's constant, $\sigma_0(l)$ is the number of positive 
divisors of $l$, $K_0$ is the modified Bessel function of the second kind, and 
$\partial_\tau = \partial /\partial \tau$. Notably, $\mathcal{S}$ obeys the 
symmetry relation $\mathcal{S}(\tau,p)=\mathcal{S}(p,\tau)$. One may show that 
this result reduces to the zero temperature expression\cite{brevik09} when 
$\tau\to 0$\cite{ellingsen10}, and we have verified that when $p=1$ the 
coefficients of the two leading-order terms as $\tau\to \infty$, of order 
$\tau$ and $\tau\ln\tau$, equal twice those found previously for the 
cylindrical shell in Ref.~\refcite{bordag02} as they should.

\section{On the  Electromagnetic Energy Momentum Tensor in Media}

The expression (\ref{3}) for the spatial part of the
energy-momentum tensor (equal to minus the Maxwell stress tensor
$T_{ik}$ according to M{\o}ller \cite{moller72}), is common for
the Minkowski and Abraham tensor alternatives as long as the medium
is isotropic. The extraction of the ``correct" form of the tensor has
however been discussed for a long time. Thus it is to be noted
that

$\bullet$  the expression is different from that of Einstein and
Laub (1908) \cite{einstein08};

$\bullet$  it is different from that of Peierls (1976)
\cite{peierls76};

$\bullet$  and it is different from that of Raabe and Welsch
(2005) \cite{raabe05}; cf. also the comment \cite{brevikPRA09} of
Brevik and Ellingsen. A survey up to 1979 is given by Brevik
\cite{brevik79}.

When combined with the Minkowski momentum density ${\bf g}^M=\bf
D\times B$ one obtains the Minkowski energy-momentum tensor
$S_{\mu\nu}^M$ whose covariant form can be written as
\cite{moller72}
\begin{equation}
S_{\mu\nu}^M=F_{\mu\alpha}H_{\nu\alpha}-\frac{1}{4}\delta_{\mu\nu}
F_{\alpha\beta}H_{\alpha\beta}
.
\end{equation}
Here $F_{4k}=iE_k, H_{4k}=iD_k, F_{ik}=B_l$ (cycl), $H_{ik}=H_l$
(cycl). This energy-momentum tensor has several attractive
properties: it has zero divergence for a pure radiation field, the
four  components of momentum and energy forming a four-vector; it
is a convenient expression for field quantization (for instance,
the quantization for a radiation field in a nondispersive medium
can be found via a mapping technique leading one from vacuum to a
medium \cite{brevik70}), and it has the peculiar property of being
a space-like expression leading to negative photon energies in
certain coordinate systems, strikingly found in connection with
the Cherenkov effect in the emitter's rest frame.

One might think that it should be relatively easy to test the
Minkowski tensor by measuring electromagnetic forces in optics.
Actually, this is not so easy as most experiments measure only the
surface force density ${\bf f}=-\frac{1}{2}E^2{\bf
\nabla}\varepsilon$, acting on surfaces. Let us give some
examples:

The classic experiment of Ashkin and Dziedzic \cite{ashkin73}
showed how a narrow light beam incident on a free liquid surface
acts by giving rise to an outward pull. A related experiment is
that of Zhang and Chang \cite{zhang88}, demonstrating the
oscillations of a water droplet by a laser pulse. The optical
stretcher experiment of Guck {\it et al.} \cite{guck01} is of the
same kind, as is the series of two-fluid experiments of Delville
{\it et al.} \cite{delville06} near the critical point. The
important factor in these experiments is simply the surface force,
 not electromagnetic momentum. And the recent
fiber experiment of She {\it et al.} \cite{she08} belongs in our
opinion to the same category \cite{brevikPRL09}. The experiment of
Campbell {\it et al.} \cite{campbell05} is however different in
nature, as it is one of the very few experiments being able to
test the Minkowski momentum directly.

Under stationary conditions, where the high-frequency forces
average out when averaged over a period, the Minkowski tensor is
able to describe all experiments that we are aware of. And this
gives support to our expression  (\ref{3}) for the stress tensor.

The work of KAM was supported by the
US National Science Foundation under Grant No.~PHY-0554926  and by the
US Department of Energy under Grants Nos.~DE-FG02-04ER41305 and 
DE-FG02-04ER-46140.


\end{document}